\documentclass[12pt,a4paper]{article}

\usepackage{geometry}
\usepackage{amssymb,amsmath}
\usepackage{natbib}
\usepackage[dvips]{graphicx}

\newcommand{\Fig}[1]{Fig~\ref{fig:#1}}
\newcommand{\Eq}[1]{Eq~(\ref{eq:#1})}
\newcommand{\Dp}[2]{\frac{\partial #1}{\partial #2}}
\newcommand{\Sec}[1]{Sec~\ref{sec:#1}}
\newcommand{\dr}{\mathrm{d}}
\newcommand{\Dr}{\mathrm{D}}
\newcommand{\er}{\mathrm{e}}
\newcommand{\UR}{\overline{u'_i u'_j}}

\title{Surface Wave Effects in the NEMO Ocean Model: Forced and Coupled
       Experiments\footnote{Final version published in \textit{J Geophys
Res: Oceans}, 2015, doi:10.1002/2014JC010565.}}

\author{{\O}yvind Breivik\thanks{Corresponding author. E-mail:
\texttt{oyvind.breivik@met.no}; ORCID Author ID: \texttt{0000-0002-2900-8458}.
Presently at the Norwegian Meteorological Institute, Alleg 70, NO-5007 Bergen, Norway}, 
   \thanks{European Centre for Medium-Range Weather
        Forecasts (ECMWF)}
   \and Kristian Mogensen\footnotemark[3]
   \and Jean-Raymond Bidlot\footnotemark[3]
   \and Magdalena Alonso Balmaseda\footnotemark[3]
   \and Peter A.E.M. Janssen\footnotemark[3]}

\begin{document}
\maketitle

\begin{abstract}
The NEMO general circulation ocean model is extended to incorporate three
physical processes related to ocean surface waves, namely the surface stress
(modified by growth and dissipation of the oceanic wave field), the turbulent
kinetic energy flux from breaking waves, and the Stokes-Coriolis force.
Experiments are done with NEMO in ocean-only (forced) mode and coupled
to the ECMWF atmospheric and wave models.  Ocean-only integrations are
forced with fields from the ERA-Interim reanalysis.  All three effects
are noticeable in the extra-tropics, but the sea-state dependent turbulent
kinetic energy flux yields by far the largest difference.  This is partly
because the control run has too vigorous deep mixing due to an empirical
mixing term in NEMO. We investigate the relation between this \textit{ad
hoc} mixing and Langmuir turbulence and find that it is much more effective
than the Langmuir parameterization used in NEMO.  The biases in sea surface
temperature as well as subsurface temperature are reduced, and the total
ocean heat content exhibits a trend closer to that observed in a recent ocean
reanalysis (ORAS4) when wave effects are included. Seasonal integrations of
the coupled atmosphere-wave-ocean model consisting of NEMO, the wave model
ECWAM and the atmospheric model of ECMWF similarly show that the sea surface
temperature biases are greatly reduced when the mixing is controlled by the
sea state and properly weighted by the thickness of the uppermost level
of the ocean model.  These wave-related physical processes were recently
implemented in the operational coupled ensemble forecast system of ECMWF.
\end{abstract}

\section{Introduction}
\label{sec:intro}
Surface waves affect the ocean surface boundary layer (OSBL) through a number
of processes, but perhaps most visibly through breaking waves which can
be seen as whitecaps on the ocean surface \citep{monahan71,wu79}. These
breaking waves enhance the turbulence in the upper part of the ocean
significantly \citep{craig94,craig96}.  Waves absorb energy and momentum
from the wind field when they grow and in turn release it when they break
\citep{janssen04,rascle06,ardhuin06,janssen12}. This lowers or raises the
stress on the water side (i.e., the stress below the oceanic wave field)
relative to the air-side stress, depending on whether the sea state is growing
or decaying.  Only when the wave field is in equilibrium with the energy
injected by the wind will the stress on the two sides of the surface be equal.

Through the interaction with the Coriolis effect, the Stokes drift
velocity associated with the wave field adds an additional term to the
momentum equation. The effect was first presented by \citet{hasselmann70}
and has since been investigated for idealized cases by \citet{weber83},
\citet{jenkins87b}, \citet{mcwilliams99} and \citet{mcwilliams00} among others.
The force is variously known as the Stokes-Coriolis force or the Hasselmann
force depending on whether it is considered to be purely an effect of the
average Coriolis force acting on a particle with a Lagrangian velocity as
given by the mean currents and the waves or as a tilting of the planetary
vorticity \citep{polton05,brostrom14}. The force does not directly modify
the total mass transport but it will alter the distribution of momentum over
the depth of the Ekman layer \citep{mcwilliams99,polton09}.

The impact of the oceanic wave field on upper-ocean mixing and mean properties
has been studied in a number of single-column mixed-layer model experiments
\citep{craig94,mcwilliams99,mcwilliams00,burchard01,kantha04,mellor04,rascle06,ardhuin06,huang11,janssen12}.
Several studies have employed large eddy simulations (LES) to
investigate the impact of Langmuir turbulence in the upper ocean
\citep{sky95,mcwilliams97,teixeira02,polton07,grant09}, and in some cases
even direct numerical simulations (DNS) have been employed \citep{sullivan04}.
Most of these studies find that waves do indeed seem to have a rather profound
impact on the upper part of the ocean, but there is still considerable
disagreement about which processes are more important.  So far there have been
few studies of the wave impact on three-dimensional ocean circulation models or
fully coupled models of the ocean, the atmosphere and the oceanic wave field
although the potential impact of waves on the climate system is recognized
\citep{babanin09b,cav12,fan14}.  \citet{fan09} demonstrated the importance
of correctly modelling momentum and energy fluxes from the wave field to the
ocean under hurricane conditions.  \citet{fan14} found that the introduction
of Langmuir turbulence following the parameterizations by \citet{mcwilliams00}
and \citet{smyth02} as well as the parameterization of mixing by non-breaking
waves suggested by \citet{qiao04} significantly changed the upper-ocean
temperature in long-term coupled climate integrations.  This latter mixing
process appears similar to the mixing due to the high Reynolds numbers
of the orbital motion of non-breaking waves explored by \citet{babanin06}
and \citet{babanin09}. Using a climate model of intermediate complexity,
\citet{babanin09b} explored three wave-related mixing processes, namely
injection of turbulent kinetic energy from breaking waves, Langmuir circulation
and the aforementioned mixing by non-breaking waves.  Like \citet{fan14}
they found that all three processes contributed to the mixed layer depth
and the temperature of the mixed layer.  Similarly, \citet{huang11} coupled
WAVEWATCH III \citep{tolman02} to a version of the Princeton Ocean Model
\citep{blu87} and demonstrated an improved summertime temperature profile
using the non-breaking parameterization of \citet{qiao04}. They found very
little direct impact by the breaking waves on the temperature.

These wave-driven processes influence the vertical structure of the
temperature and current fields in the mixed layer in general, and in the
upper few meters in particular. This has implications for coupled models as
these processes will affect the feedback between the ocean and the atmosphere
\citep{janssen13}.  However, on shorter time scales and at higher spatial
resolution it is also clear that these processes will influence the drift of
objects and pollutants on the sea surface or partially or wholly submerged.
This has practical importance for oil spill modelling \citep{hac06}, search
and rescue \citep{bre08,dav09,bre13} and dispersion of biological material
\citep{rohrs14}.

The NEMO ocean model \citep{madec12} has been coupled to the atmospheric model
with wave forcing from the wave model as part of the ensemble suite of the
Integrated Forecast System (IFS) of the European Centre for Medium-Range
Weather Forecasts (ECMWF) since November 2013 (IFS Cycle 40R1). Here we
describe the implementation of the three wave effects mentioned above in
forced (ocean-only) integrations of NEMO using forcing from the ERA-Interim
reanalysis \citep{dee11} as well as their implementation in a fully coupled
atmosphere-wave-ocean seasonal forecast system.  

The paper is organized as follows. \Sec{theory} describes the processes that
have been implemented and lays out their actual implementation in NEMO.
\Sec{forced} describes the results of long ocean-only integrations and
compares with control runs, observations and the ORAS4 ocean reanalysis
\citep{balmaseda13}. \Sec{coupled} describes the coupled atmosphere-wave-ocean
coupling used for seasonal integrations and compares the results of a control
run where no direct coupling exists between the wave model and the ocean model
to a run where NEMO is forced with stresses, turbulent fluxes and Stokes
drift from the wave model ECWAM \citep{wam40r1}. \Sec{disc} discusses the
results and the deficiencies in the existing model setup. \Sec{conc}
concludes and makes suggestions for further work on the investigation of wave
effects in ocean-only as well as coupled atmosphere-wave-ocean models.

\section{Wave effects in the Ocean Surface Boundary Layer}
\label{sec:theory}
Introducing wave forcing in an Eulerian ocean model entails communicating
the relevant forcing fields from a wave model. We start with a brief
presentation of spectral wave models and how the two-dimensional wave spectrum
relates to the fluxes and fields that have a bearing on the ocean surface
boundary layer.

\subsection{Fluxes and fields estimated from a spectral wave model}
\label{sec:spectral}
Third generation spectral wave models
\citep{has88,tol91,kom94,ris99,jan04,tolman02,holthuijsen07,cav07} solve
the action balance equation (see Eq (1.185) by \citet{kom94} and Eq (2.71)
by \citet{jan04}) for the wave action density $N$ (a function of the
Cartesian co-ordinate $\mathbf{x}$, frequency $f$ and direction $\theta$) as
follows,
\begin{equation}
 \left(\Dp{}{t} + \Dp{\omega}{\mathbf{k}} \cdot
 \Dp{}{\mathbf{x}} - \Dp{\omega}{\mathbf{x}} \cdot \Dp{}{\mathbf{k}}\right) N 
 = S'_\mathrm{in} + S'_\mathrm{nl} + S'_\mathrm{ds}.
 \label{eq:actionbal}
\end{equation}
Here, the right-hand source terms refer to wind input (in), nonlinear transfer
(nl), and dissipation due to wave breaking (ds), respectively. The dissipation
term may include shallow-water effects and bottom friction.
The gradient in frequency represents
shoaling and refraction, and $\omega = \sigma + \mathbf{k}\cdot \mathbf{u}$ is
the absolute frequency as seen by an observer standing still, whereas $\sigma$
is the intrinsic frequency as seen by an observer moving with the current.  We
also note that
\begin{equation}
   \Dp{\omega}{\mathbf{k}} \equiv \mathbf{c}_\mathrm{g} + \mathbf{u},
 \label{eq:doppler}
\end{equation}
where $\mathbf{c}_\mathrm{g}\,$ is the wave group veloctiy vector.
The wave action density is related to the wave variance density through
$N = F/\sigma$.
In deep water with no current refraction
\Eq{actionbal} reduces to the energy balance equation,
\begin{equation}
 \Dp{F}{t} + \nabla \cdot (\mathbf{c}_\mathrm{g} F) 
 = S_\mathrm{in} + S_\mathrm{nl} + S_\mathrm{ds},
 \label{eq:energybal}
\end{equation}
written here in flux form. Note that the source terms in \Eq{energybal}
are related to those in the action balance equation (\ref{eq:actionbal})
as $S = \sigma S'$.

\subsection{The air-side stress modified by surface waves}
\label{sec:cdww}
The presence of an undulating surface affects the roughness felt by the
airflow.  The atmospheric momentum flux to the oceanic wave field is denoted
$\tau_\mathrm{in}$. It is convenient to define an air-side friction velocity
in relation to the total air-side stress, $\tau_\mathrm{a}$, as
\begin{equation}
   u_*^2 = \tau_\mathrm{a}/\rho_\mathrm{a}.
\end{equation}
Here, $\rho_\mathrm{a}$ is the surface air density.
\citet{charnock55} was the
first to relate the roughness of the sea surface to the friction velocity,
\begin{equation}
   z_0 = \alpha_\mathrm{CH} \frac{u_{*}^2}{g},
\end{equation}
where $\alpha_\mathrm{CH}$ is known as the Charnock constant.
\citet{janssen89,janssen91} assumed that $\alpha_\mathrm{CH}$ is not constant
but varies with the sea state,
\begin{equation}
   \alpha_\mathrm{CH} = 
   \frac{\hat{\alpha}_\mathrm{CH}}
        {\sqrt{1-\tau_\mathrm{in}/\tau_\mathrm{a}}},
   \label{eq:modcharnock}
\end{equation}
where $\hat{\alpha}_\mathrm{CH} = 0.006$ \citep{bidlot12} and the wave-induced
stress, $\tau_\mathrm{in}$, is related to the wind input to the wave field as
\begin{equation}
   \boldsymbol{\tau}_\mathrm{in} = \rho_\mathrm{w}g \int_0^{2\pi} \!
      \int_0^{\infty} \frac{\mathbf{k}}{\omega} S_\mathrm{in} \, 
      d\omega \, d\theta.
   \label{eq:tauin}
\end{equation}
Here $\rho_\mathrm{w}$ is the water density.
A diagnostic spectral tail proportional to $\omega^{-5}$ has been applied
above a cutoff frequency $\omega_\mathrm{w}$ \citep{kom94,wam40r1}.
The wave-modified drag coefficient is then
\begin{equation}
   C_\mathrm{D} = \frac{\kappa^2}{\log^2(10/z_0)},
\end{equation}
where $\kappa = 0.4$ is von K\'{a}rm\'{a}n's constant.  Note that the drag
coefficient as defined here is related to the 10-m \textit{neutral} wind speed,
$U_\mathrm{10N}$. This drag coefficient is computed by ECWAM.  The Charnock
parameter (\ref{eq:modcharnock}) is the main coupling mechanism between the
atmosphere and the wave field in IFS, in place since 1998 \citep{jan04}.

\subsection{The water-side stress modified by surface waves}
\label{sec:tauoc}
As the wind increases, the wave field responds by first growing and storing
more momentum. In this phase there is a net influx of momentum to the wave
field. Then, as the waves mature and the breaking intensifies, the momentum
flux from the wave field to the ocean starts to close on the flux from the
atmosphere to the waves. This is the equilibrium state where dissipation
matches wind input, also referred to as fully developed windsea (since the
waves cannot become higher), see e.g. \citet{kom94,wmo98,holthuijsen07}.
Finally, as the wind dies down there will be a net outflux of momentum from
the wave field, almost all of which will go to the ocean.

If wind input and dissipation in the wave field were in equilibrium, the
air-side stress would be equal to the total water-side stress. By water-side
stress is meant the stress as seen by the Eulerian ocean, i.e., the momentum
flux from the waves. However, most of the time waves are not in equilibrium
\citep{janssen12,janssen13}, giving differences in air-side and water-side
stress of the order of 5--10\%, with occasional departures much larger in
cases where the wind suddenly slackens. Likewise, in cases with sudden onset
of strong winds the input from the wind field will be much larger than the
dissipation to the ocean, lowering the water-side stress to values well below
70\% of its normal ratio to the air-side stress.  The water-side stress thus
equals the total atmospheric stress minus the momentum flux absorbed by the
wave field (positive) minus the momentum injected from breaking waves to the
ocean (negative), $\boldsymbol{\tau}_\mathrm{oc} = \boldsymbol{\tau}_\mathrm{a}
- \boldsymbol{\tau}_\mathrm{in} - \boldsymbol{\tau}_\mathrm{ds}$. Here,
the dissipation source term is assumed to include all relevant dissipative
processes. This can be written \citep{wam40r1}
\begin{equation}
   \boldsymbol{\tau}_\mathrm{oc} = \boldsymbol{\tau}_\mathrm{a} -
   \rho_\mathrm{w}g \int_0^{2\pi} \! \int_0^{\infty} 
   \frac{\mathbf{k}}{\omega}(S_\mathrm{in} + S_\mathrm{ds})\,
       \dr\omega \dr\theta.
   \label{eq:tauoc}
\end{equation}
The stress from waves is archived as a normalized quantity and is applied
as a factor to the air-side stress in our implementation in NEMO.

\subsection{Mixing parameterizations}
\label{sec:wavetke}
The TKE equation with Reynolds averages can be written
\begin{equation}
     \frac{\Dr e}{\Dr t} =
      \frac{g}{\rho_\mathrm{w}}\overline{u'_3 \rho'} 
     -\UR \Dp{\overline{u}_i}{x_j} 
     -\Dp{}{x_j}(\overline{u'_je})
     -\frac{1}{\rho_\mathrm{w}}\Dp{}{x_i}(\overline{u'_ip'})
     -\epsilon.
     \label{eq:tker}
\end{equation}
Here, $e \equiv \overline{q^2}/2 = \overline{u'_i u'_i}/2$ is the TKE per
unit mass (with $q$ the turbulent velocity) and $\epsilon$ the dissipation
rate (see e.g. \citet{stu88}, p 152).  NEMO has the option of modelling the
evolution of TKE with local closure (a prognostic equation in $e$ only,
see \citealt{stu88} pp 203--208 and \citealt{pope00} pp 369--373). Assuming
that the advective terms are small in comparison and making the gradient
transport approximation where turbulent coefficients are proportional to
the gradients in the mean quantities, we arrive at
\begin{equation}
   \Dp{e}{t} = K_\mathrm{m} S^2 
             - K_\rho N^2 
             + \Dp{}{z}(K_q\Dp{e}{z})
             - c_\epsilon \frac{e^{3/2}}{l_\epsilon}.
    \label{eq:tke2}
\end{equation}
This is the standard one-equation formulation for NEMO (see the reference
manual for NEMO v3.4, \citealt{madec12}, pp 176--177). Here $l_\epsilon$
is the mixing length. The buoyancy term is assumed proportional to the local
Brunt-V\"{a}is\"{a}l\"{a} frequency,
\begin{equation}
   N^2 = -\frac{g}{\overline{\rho}}\Dp{\rho_\mathrm{w}}{z},
    \label{eq:n2}
\end{equation}
and the shear production is related to the shear of the mean flow,
\begin{equation}
   S^2 = \left(\Dp{\overline{\mathbf{u}}}{z}\right)^2.
    \label{eq:s2}
\end{equation}
Finally, the mixing length is given by a relation by \citet{blanke93},
see also \citet{gaspar90} and \citet{madec12}, pp 177--179.

Two non-standard mixing processes present in NEMO's TKE scheme warrant our
attention.  The first is an artificial boost to the TKE known as the ETAU
parameterization which is pegged to the surface TKE with an exponential
vertical decay (see \citet{madec12}, Sec 10.1),
\begin{equation}
 e_\tau(z) = 0.05 e_1 \exp z/h_\tau.
 \label{eq:etau}
\end{equation}
The depth scale $h_\tau$ can vary with longitude from 0.5~m at the Equator
to 30~m poleward of $44^\circ$ or be fixed at 10~m. The coefficient, here
$0.05$, can also be varied.  The second mixing process of interest to us
is a parameterization of Langmuir turbulence according to \citet{axell02}
which has been implemented in NEMO.  The vertical velocity $w_\mathrm{LC}$
of the Langmuir cells is assumed to peak at $H_\mathrm{LC}/2$, half the
maximum depth to which Langmuir cells penetrate,
\begin{equation}
   w_\mathrm{LC}(z) = c_\mathrm{LC} v_\mathrm{s}\sin \left(-\frac{\pi
   z}{H_\mathrm{LC}}\right), 0 > z \geq H_\mathrm{LC}.
   \label{eq:wlc}
\end{equation}
Here $v_\mathrm{s}$ is the surface Stokes drift speed and $c_\mathrm{LC}
= 0.15$ is a coefficient.  \citet{axell02}, by making an analogy with the
characteristic convective velocity scale \citep{dalessio98} further assumed the
Langmuir production term in the TKE equation (\ref{eq:tker}) could be written
\begin{equation}
   P_\mathrm{LC}(z) = \frac{w_\mathrm{LC}^3}{H_\mathrm{LC}}.
   \label{eq:lc}
\end{equation}
This production term will attain a maximum value in the interior of the
mixed layer.

\citet{craig94}, CB94 hereafter, demonstrated that as waves break they will
considerably modify the vertical dissipation profile from the traditional
law-of-the-wall where dissipation $\propto z^{-1}$ \citep{stu88}.
With wave breaking CB94 found dissipation $\propto z^{-3.4}$. \citet{ter96}
and \citet{drennan96} later demonstrated that the observed dissipation rates
under breaking waves are indeed much higher than anticipated by the law
of the wall. CB94's model has since been extended to a two-equation turbulence
model by \citet{burchard01}, who demonstrated that the injection of turbulent
kinetic energy from breaking waves was sufficient to successfully model
the evolution of the mixed layer representative of North Sea conditions.
CB94 suggested that the flux of turbulence kinetic energy (TKE) should be
related to the water friction velocity $w_*$ as
\begin{equation}
   \Phi_\mathrm{oc} = \rho_\mathrm{w} \alpha_\mathrm{CB} w_*^3.
   \label{eq:alphacb}
\end{equation}
CB94 assumed that $\alpha_\mathrm{CB}$ was a constant $\sim100$, but noted
that its range would probably be between 50 and 150, depending on the
sea state (see also \citealt{mellor04}). The TKE flux from breaking waves
is related (see e.g.  \citealt{janssen04,rascle06,janssen12,janssen13})
to the dissipation source function of a spectral wave model as
\begin{equation}
   \Phi_\mathrm{oc} = -\rho_\mathrm{w}g \int_0^{2\pi} \!
                      \int_0^{\infty} \! S_\mathrm{ds}\, 
                        \dr\omega \dr\theta 
                    = -\rho_\mathrm{a} m u_*^3.
   \label{eq:phioc}
\end{equation}
For consistency we have written the energy flux \textit{from} the waves
(thus always negative), $m \approx -\sqrt{\rho_\mathrm{a}/\rho_\mathrm{w}}
\alpha_\mathrm{CB}$, normalized by the air friction velocity $u_*$.  In NEMO
the energy flux from breaking waves is introduced as a Dirichlet boundary
condition on TKE, following \citet{mellor04}. It is assumed that in the
wave-affected layer the mixing length can be set to a constant $l_\mathrm{w}
= \kappa z_\mathrm{w}$ where the surface roughness length relates to the
significant wave height $H_\mathrm{s}$ as $z_\mathrm{w} = 0.5H_\mathrm{s}$,
and that in this near-surface region diffusion balances dissipation. In
this case the TKE equation takes a simple exponential solution [see Eq~(10)
by \citealt{mellor04}],
\begin{equation}
  e(z) = e_0 \exp(2\lambda z/3).
   \label{eq:phimb}
\end{equation}
Here the inverse length scale is
\begin{equation}
   \lambda = [3/(S_q B \kappa^2)]^{1/2}z_\mathrm{w}^{-1},
   \label{eq:lambda}
\end{equation}
with $S_q=0.2$ and $B=16.6$ given by \citet{mel82}.  This is how the flux
from breaking waves is implemented in NEMO v3.4.  This allows the following
simple boundary condition
\begin{equation}
e_0 = \frac{1}{2}\,\left(15.8\,\alpha_\mathrm{CB} \right)^{2/3}
         \,\frac{|\boldsymbol{\tau}_\mathrm{oc}|}{\rho_\mathrm{w}}.
\label{eq:esbc}
\end{equation}

However, the inverse depth scale (\ref{eq:lambda}) is sea state dependent,
and for a wave height of, say, 2.5 m, which is close to the global mean,
$\lambda^{-1} \approx 0.5\, \mathrm{m}$.  Thus, $e(z)$ varies rapidly with
depth, and we have modified the boundary condition (\ref{eq:esbc}) by weighting
the surface value by the thickness of the topmost level to attain an average
value more representative of the turbulence near the surface of the model,
\begin{equation}
   e_1 = \frac{1}{L} \int_{-L}^{0} \! e(z) \,\dr z.
   \label{eq:eavg}
\end{equation}
Here $L = \Delta z_1/2$ is the depth of the $T$-point of the first level.
Integrating \Eq{phimb} is straightforward, and the average TKE boundary
condition (\ref{eq:eavg}) becomes
\begin{equation}
  e_1 = e_0 \frac{3}{2\lambda L} \left[1 -
                       \exp(-2\lambda L/3)\right].
  \label{eq:e1}
\end{equation}
It is clear that as the vertical resolution increases the difference between
$e_1$ and $e_0$ becomes smaller, and in the limit the two coincide. The
weighting (\ref{eq:e1}) is thus less important with higher vertical resolution.
It is worth noting that the exponential profile (\ref{eq:phimb}) assumed by
\citet{mellor04} is only valid very near the surface, and in fact CB94 had
already found the solution to the more general case where the mixing length
is allowed to vary with depth.  We have not implemented this operationally,
but preliminary tests suggest that the effect is to roughly double the depth
over which the TKE from breaking waves is distributed. The derivation is
presented in the appendix.

\subsection{The Stokes-Coriolis forcing}
\label{sec:stcor}
Waves set up a Lagrangian displacement $\mathbf{v}_\mathrm{s}$ in the down-wave
direction known as the Stokes drift velocity \citep{sto47}. Although it
decays rapidly with depth, it can be substantial near the surface
($|\mathbf{v}_\mathrm{s}| \sim 0.7\, \mathrm{m\,s}^{-1}$). In combination
with the Earth's rotation it adds an additional veering to the upper-ocean
currents known as the Stokes-Coriolis force \citep{hasselmann70},
\begin{equation}
     \frac{\Dr \mathbf{u}}{\Dr t} = -\frac{1}{\rho} \nabla p 
  + (\mathbf{u} + \mathbf{v}_\mathrm{s}) \times f\hat{\mathbf{z}}
  + \frac{1}{\rho} \frac{\partial \boldsymbol{\tau}}{\partial z}.
  \label{eq:stcor}
\end{equation}
Here $f$ is the Coriolis frequency, $\hat{\mathbf{z}}$ is the upward
unit vector, $p$ is the pressure and $\boldsymbol{\tau}$ is the stress.
The full two-dimensional spectrum is in principle required to compute the
Stokes drift velocity profile \citep{janssen04,janssen12},
\begin{equation}
   \mathbf{v}_\mathrm{s}(z) = 4\pi \int_0^{2\pi}\!\int_0^{\infty} 
                              f \mathbf{k} \er^{2kz} F(f,\theta) \, 
                              \dr f\, \dr\theta.
   \label{eq:uvfth}
\end{equation}
This is computationally demanding and full two-dimensional wave spectra from a
numerical wave model (see e.g. \citealt{wam40r1}) may not always available. It
is therefore customary to introduce a simplified, monochromatic Stokes drift
profile (see e.g., \citet{car05,polton05,saetra07,tamura12}). However, it
was shown by \citet{bre14} that this profile is a poor match to the
full profile and that the following parameterization gives a considerable
improvement,
\begin{equation}
   \mathbf{v}_\mathrm{e}(z) = \mathbf{v}_0 
   \frac{\er^{2k_\mathrm{e}z}}{1-8k_\mathrm{e}z}.
   \label{eq:uve1}
\end{equation}
Here the subscript ``e'' distinguishes the approximate profile from the full
Stokes drift velocity profile (\ref{eq:uvfth}). The surface Stokes drift
velocity vector $\mathbf{v}_0$ is computed by ECWAM and is available both
in ERA-Interim \citep{dee11} and from the operational ECMWF forecasts 
\citep{wam40r1}.

To compute the profile (\ref{eq:uve1}) we must find the inverse depth scale
$k_\mathrm{e}$. This is related to the transport $T_\mathrm{s}$ through
the exponential integral $E_1$ \citep{abr72} and can be solved analytically
\citep{bre14} to yield
\begin{equation}
   T_\mathrm{s} = \frac{|\mathbf{v}_0| \er^{1/4} E_1(1/4)}{8 k_\mathrm{e}}.
   \label{eq:UVe} 
\end{equation}
Rearranging we get the following expression for the inverse depth scale,
\begin{equation}
   k_\mathrm{e} = \frac{|\mathbf{v}_0| \er^{1/4} E_1(1/4)}{8 {T}_\mathrm{s}}.
   \label{eq:ke} 
\end{equation}
Here $E_1(1/4) \approx 1.34$, thus
\begin{equation}
   k_\mathrm{e} \approx \frac{|\mathbf{v}_0|}{5.97{T}_\mathrm{s}}.
   \label{eq:keapprox} 
\end{equation}
The $n$-th order spectral moment is defined as
\begin{equation}
   m_{n} = \int_0^{2\pi} \! \int_0^{\infty} 
           f^{n} F(f,\theta) \, \dr f\, \dr\theta.
   \label{eq:moment}
\end{equation}
The mean frequency is defined as $\overline{f} = m_1/m_0$
\citep{wmo98,holthuijsen07} and the significant wave height $H_\mathrm{s} =
4\sqrt{m_0}$.  We can derive the first moment from the integrated parameters
of a wave model or from wave observations and find an estimate for the
Stokes transport,
\begin{equation}
  \mathbf{T}_\mathrm{s} \approx \frac{2\pi}{16} \overline{f} H_{m_0}^2
  \hat{\mathbf{k}}_\mathrm{s}.
  \label{eq:UVHsf}
\end{equation}
Here $\hat{\mathbf{k}}_\mathrm{s} = (\sin \theta_\mathrm{s},
\cos \theta_\mathrm{s})$ is the unit vector in the direction
$\theta_\mathrm{s}$ of the Stokes transport. We approximate the Stokes
transport direction by the surface Stokes drift (see \citet{bre14}).

\section{Ocean-only Forced Model Experiments}
\label{sec:forced}
The NEMO model is run on a tripolar ORCA~1$^\circ$ grid configuration
with 42 vertical levels. The uppermost level is 10~m thick. The model
is coupled to LIM2, a two-level thermodynamic-dynamic sea ice model
\citep{fichefet97,bouillon09} and is relaxed weakly towards a climatology in
temperature (3-yr e-folding time). No sea surface temperature (SST) relaxation
is performed. The ORCA grid is such that the resolution is increased towards
the Equator (roughly $1/3^\circ$) to better resolve tropical waves (see
\citet{madec12} for details on the ORCA grid).

The atmospheric and wave forcing fields have been computed from the ERA-Interim
reanalysis \citep{simmons07,dee11}.  ERA-Interim is a continuously updated
atmospheric and wave field reanalysis starting in 1979.  The resolution
of the wave model is $1.0^\circ$ on the Equator but the resolution is
kept approximately constant globally through the use of a quasi-regular
latitude-longitude grid where grid points are progressively removed toward the
poles \citep{jan04}. Similarly, the atmospheric model fields are archived
on a reduced Gaussian grid of approximately $0.75^\circ$ resolution at
the Equator.  Some care has to be taken when interpolating between these
grids, in particular where wave parameters are interpolated from the
ECWAM grid to the ORCA grid. NEMO requires fluxes to be defined in all ocean
points. However, there are discrepancies between the ice coverage and the
land-sea mask of the wave grid and the ocean grid. This is solved by reverting
to the ECMWF drag law \citep{janssen08,edson13} where ECWAM has ice or land,
\begin{equation}
   C_\mathrm{D}(z=10 \, \mathrm{m}) = \left(a +
     bU_{10}^{p_1}\right)/U_{10}^{p_2}.
   \label{eq:cdec}
\end{equation}
The coefficients are $a = 1.03 \times 10^{-3}$, $b = 0.04\times 10^{-3}$,
$p_1 = 1.48$ and $p_2 = 0.21$. Here $U_{10}$ is the 10-m wind speed from
ERA-Interim. Where ECWAM and NEMO agree on open water, the stress is computed
from the drag coefficient of ECWAM,
\begin{equation}
   \tau_\mathrm{a} = \rho_\mathrm{a}C_\mathrm{DW}U^2_{10\mathrm{N}}.
   \label{eq:taua}
\end{equation}
Here, $U_{10\mathrm{N}}$ is the neutral 10-m wind speed, available on the
ECWAM grid.  The conversion to water-side stress is implemented as
\begin{equation}
   \tau_\mathrm{oc} = \tilde{\tau}\tau_\mathrm{a},
   \label{eq:tautilde}
\end{equation}
where $\tilde{\tau}$ is the ratio of water-side to air-side stress (see also
Eq~\ref{eq:tautilde}). It is this parameter which is archived by ERA-Interim.

A standard integration period covering the ERA-Interim period from 1979
up until the end of 2009 has been used in the following. A summary of the
settings for the model runs can be found in Tables~\ref{tab:commonsettings}
and~\ref{tab:settings}. Four experiments with the new wave-related effects
are presented, all compared against a control experiment, \textbf{CTRL},
where standard settings are used for NEMO.  The CTRL experiment includes a
parameterization of the TKE flux from breaking waves (CB94) without explicit
sea state information (see \Eq{alphacb} and also \citet{madec12}, Sec 10.1)
but has no averaging over the topmost model level (\ref{eq:eavg}).  The stress
in CTRL is computed using the ECMWF drag law (\ref{eq:cdec}) with air-side
stress (\ref{eq:taua}).  In all runs the NEMOVAR observation operator
$\mathbf{y} = \mathcal{H}(\mathbf{x})$ relating model state $\mathbf{x}$
to observation space ($\mathbf{y}$) is applied.  This allows a comparison
of model integrations (in our case without assimilation) against the large
number of quality-controlled temperature and temperature-salinity profiles
compiled in the EN3 data set \citep{ingleby07}.  Seasonal averages (December
to January, DJF, and June to August, JJA) over the period 1989-2008 are used
to compare SST fields in the following.

\begin{figure}[h]
\begin{center}
\includegraphics[scale=1.0]{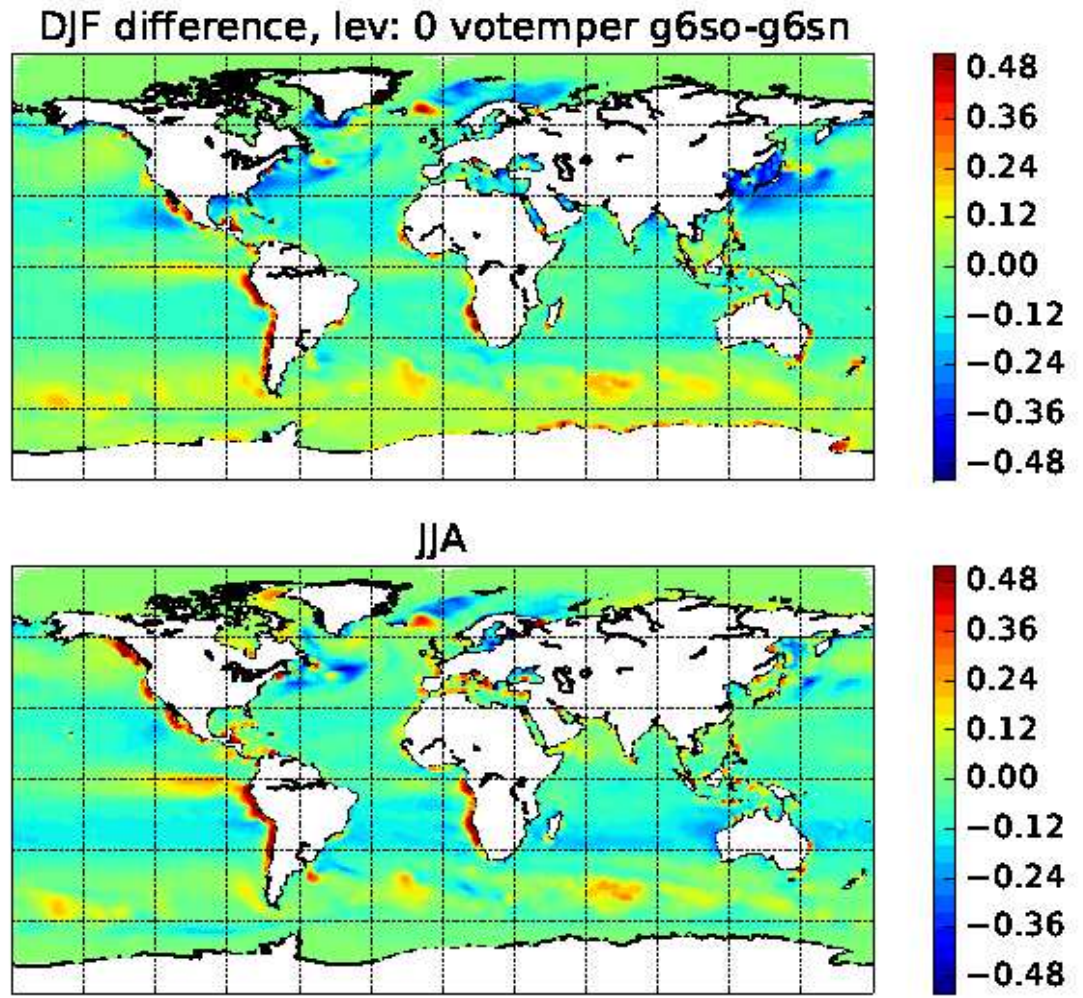}\\
\caption{Long-term SST differences between a run with water-side stress modulated by
the ECWAM wave model (TAUOC experiment, see Table 2) and the CTRL run.}
\label{fig:tauoc-ctrl}
\end{center}
\end{figure}

The first wave experiment, \textbf{TAUOC}, uses the water-side stress described
in \Sec{tauoc} together with the ECWAM drag coefficient (see \Sec{cdww}).
The effect is confined mostly to areas with rapidly developing weather systems
in the extra-tropics (\Fig{tauoc-ctrl}), where the sea state will be quite
far from equilibrium.  There is a slight weakening of the wind stress along
the west coast of South America and along the coast of south-west Africa,
leading to decreased upwelling.  This is mainly a consequence of differences
between the ECWAM drag coefficient and the drag law (\ref{eq:cdec}) caused by
limited fetch near the coast. The overall effect is a slight reduction of the
bias in the tropics compared to EN3 near-surface temperature measurements
(0.1 K, not shown).  In the tropics, differences are most likely due to
differences in the drag over swell compared to the drag law (\ref{eq:cdec})
used for the CTRL experiment.

\begin{figure}[h]
\begin{center}
\includegraphics[scale=1.0]{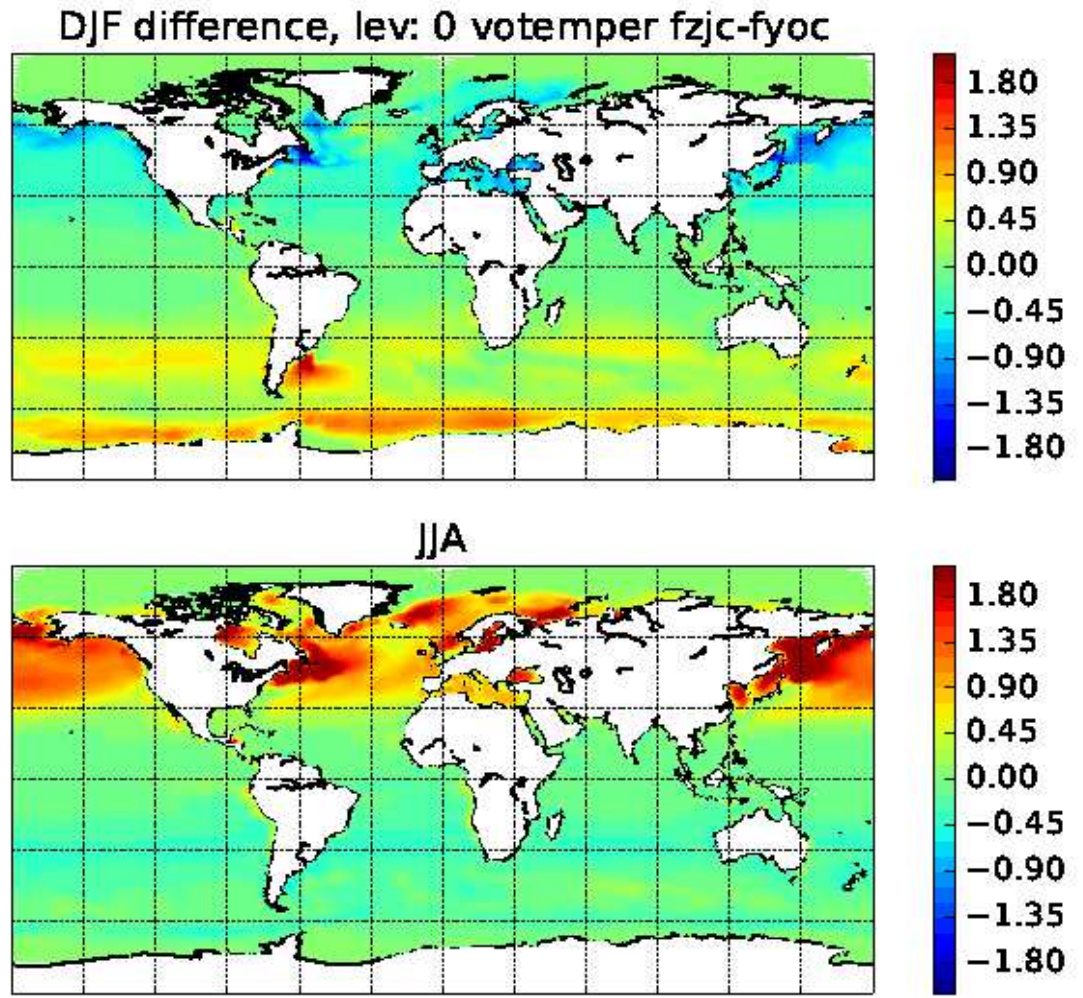}\\
\caption{Long-term SST differences between a run with TKE flux from 
ECWAM (WTKE experiment, see Table 2) and the CTRL run. 
Note that the color scale is $\pm 2$ K.}
\label{fig:wtke-ctrl}
\end{center}
\end{figure}

The second experiment, \textbf{WTKE}, introduces the TKE flux (\ref{eq:phioc})
from ECWAM (described in \Sec{wavetke}). The differences found in the
extra-tropics amount to more than 2~K (\Fig{wtke-ctrl}). The large difference
suggests that the standard settings of NEMO overdoes the mixing due to waves,
especially with coarse vertical resolution (cf Eq~\ref{eq:eavg}). 

\begin{figure}[h]
\begin{center}
\includegraphics[scale=1.0]{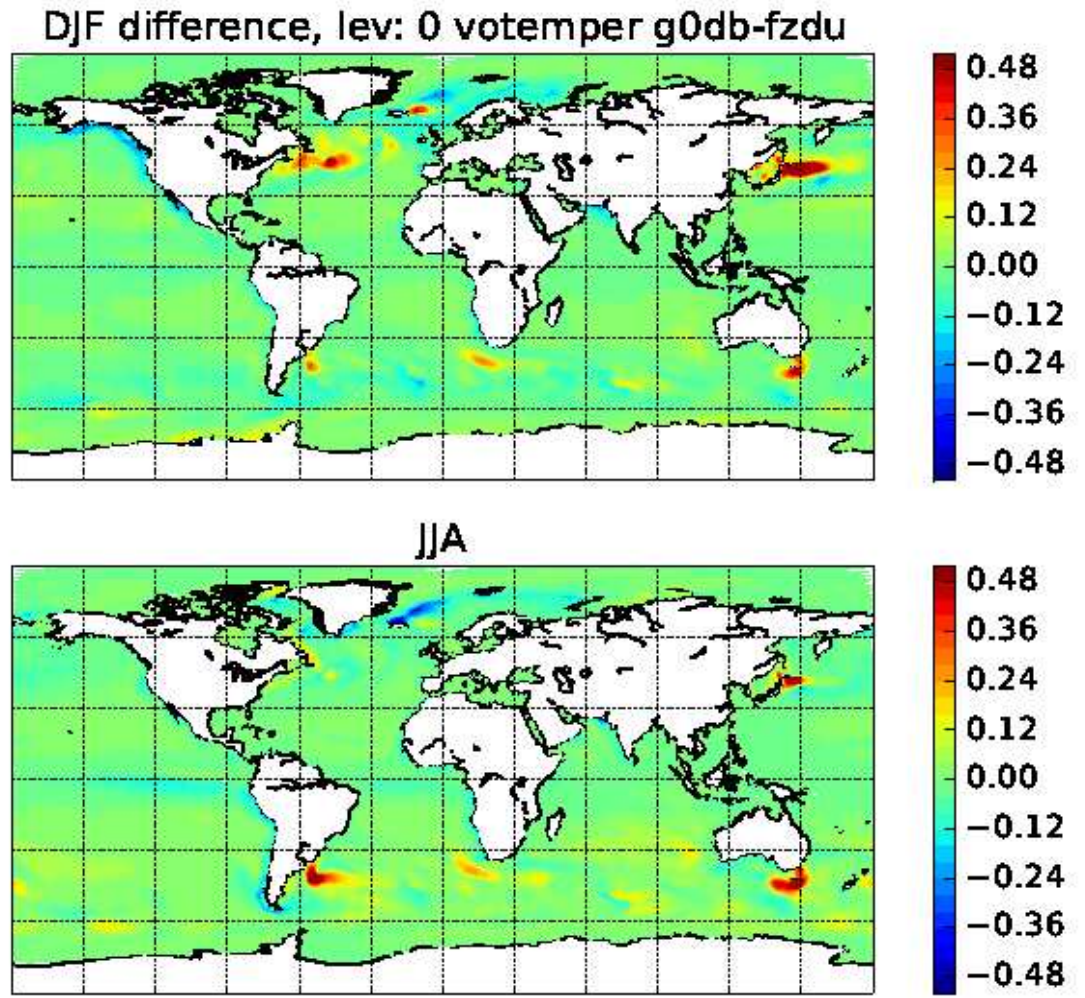}\\
\caption{Long-term SST differences between a run with Stokes-Coriolis forcing (STCOR experiment,
see Table 2) and the CTRL run.}
\label{fig:stcor-ctrl}
\end{center}
\end{figure}

The \textbf{STCOR} experiment introduces Stokes-Coriolis forcing from ECWAM
as described in \Sec{stcor}.  The largest impact (\Fig{stcor-ctrl}) is found
in areas with extra-tropical cyclones in combination with strong temperature
gradients, such as across the Gulf Stream and the Kuro-Shio.

Another experiment called \textbf{LOW} where a Law-of-the-wall boundary
condition is applied, i.e. no TKE flux from breaking waves, was
also performed.  The motivation is to establish a lower bound on the mixing.
The difference between the WTKE and LOW runs is much smaller (not shown)
than the difference between CTRL and WTKE.

\begin{figure}[h]
\begin{center}
\begin{tabular}{cc}
   (a) & \includegraphics[scale=0.55]{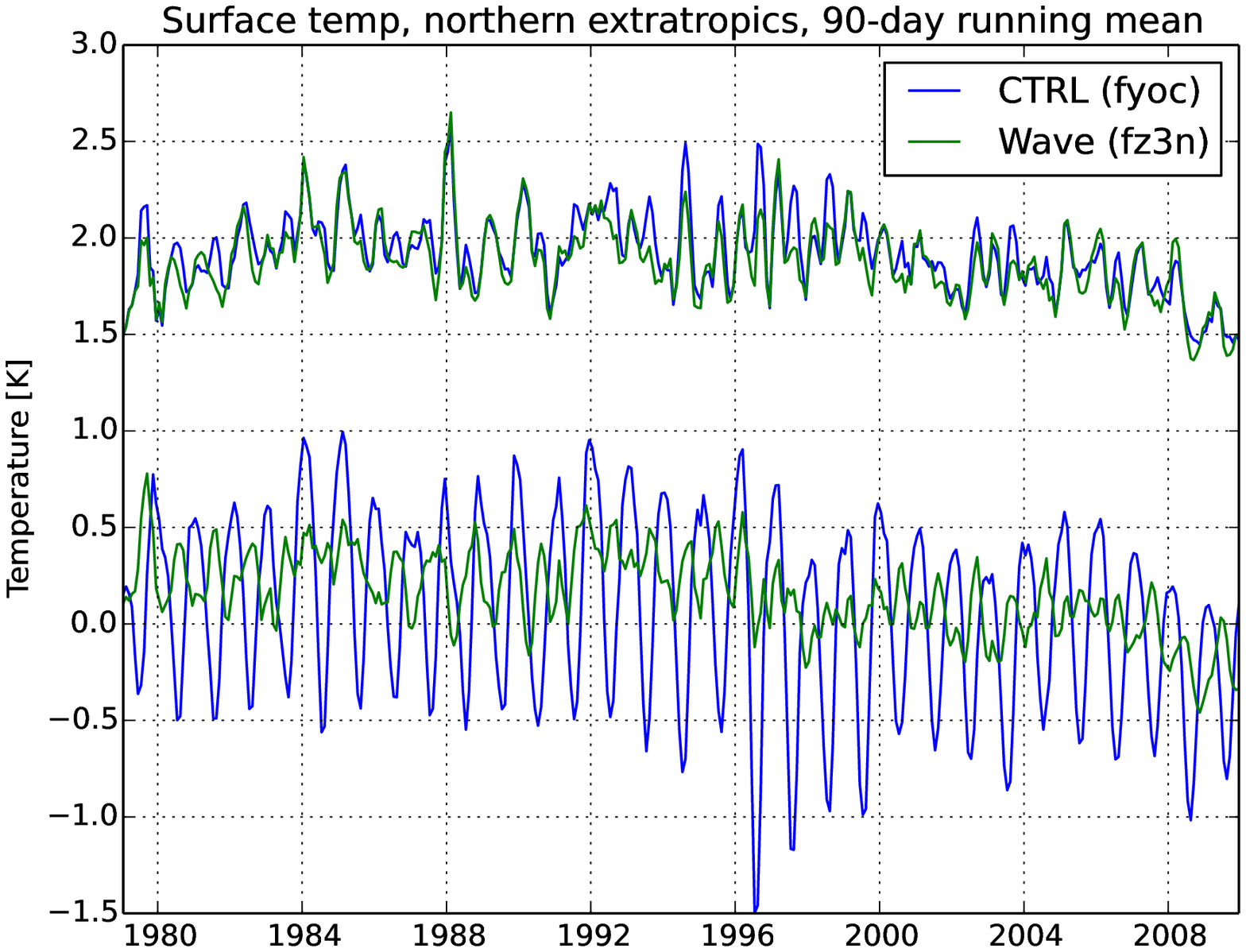}\\
   (b) & \includegraphics[scale=0.55]{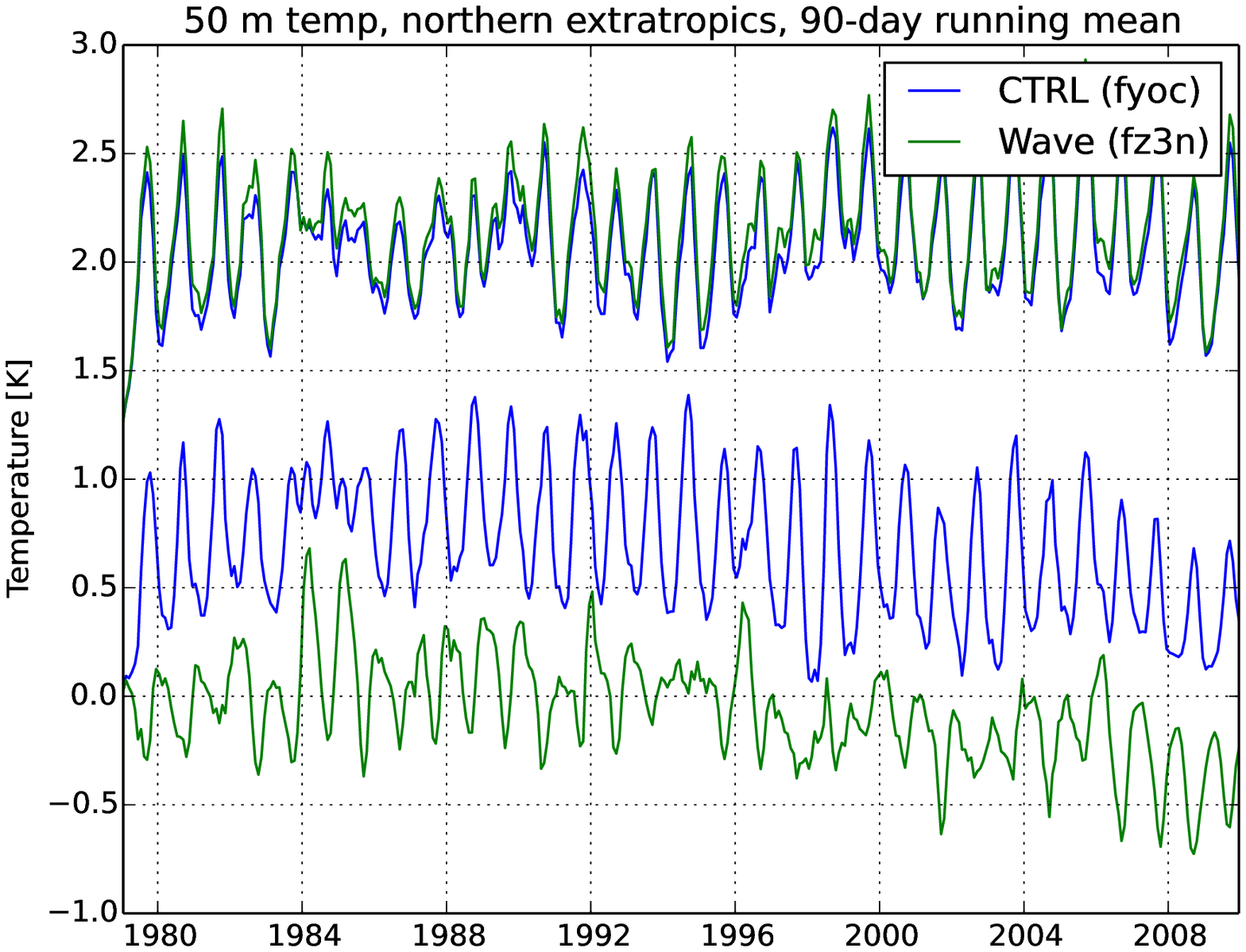}\\
\end{tabular}
\caption{A comparison of surface (a) and 50 m depth (b) EN3 temperature
observations \citep{ingleby07} in the northern extra-tropics in the CTRL run
(blue) and a run with all three wave effects switched on (green), see Table 2
for more information about the experiments. The upper curves show the standard
deviation while the lower curves represent the bias. A 90-day running mean is
employed.}
\label{fig:nxtrp}
\end{center}
\end{figure}

\begin{figure}[h]
\begin{center}
\begin{tabular}{cc}
 (a) & \includegraphics[scale=0.55]{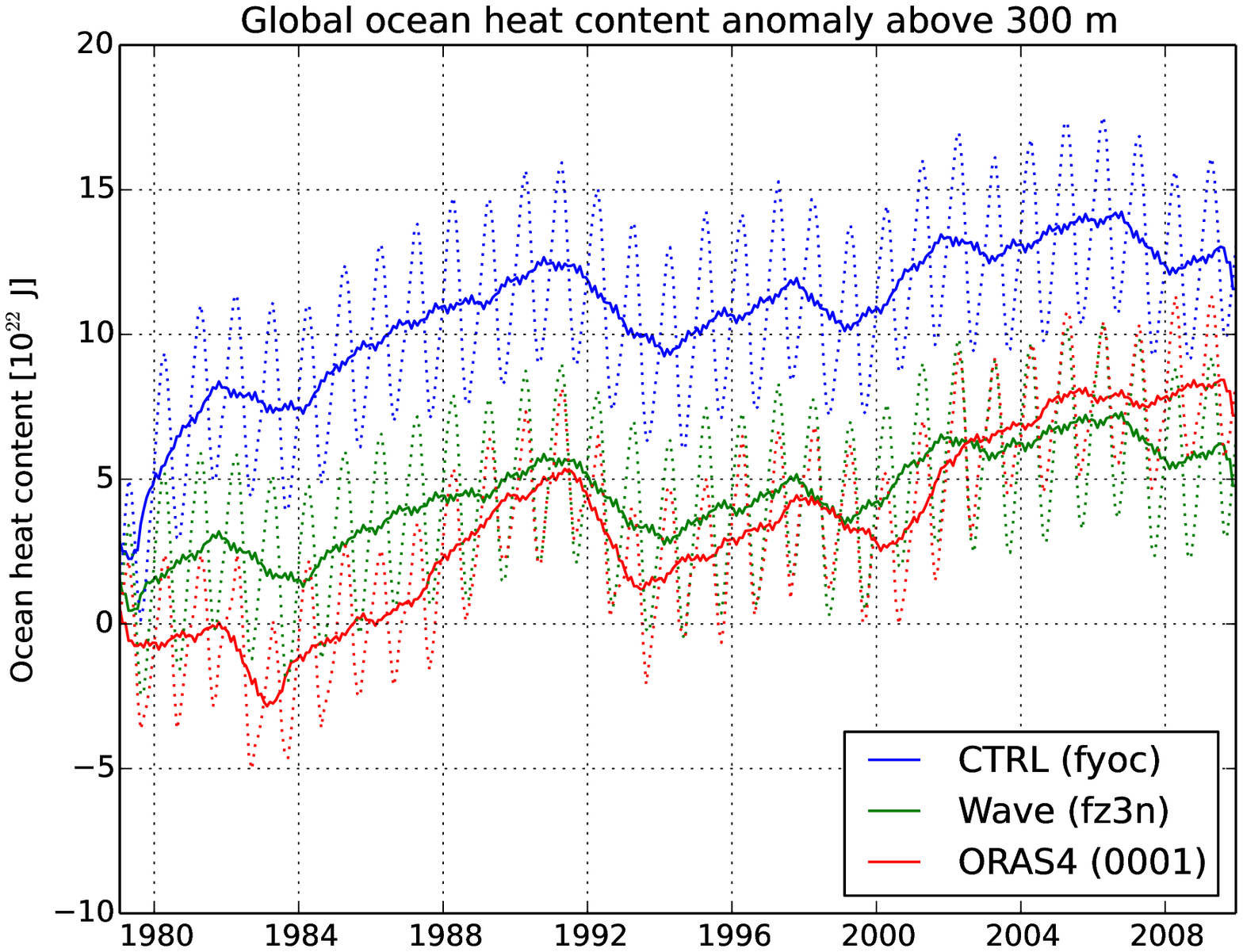}\\
 (b) & \includegraphics[scale=0.55]{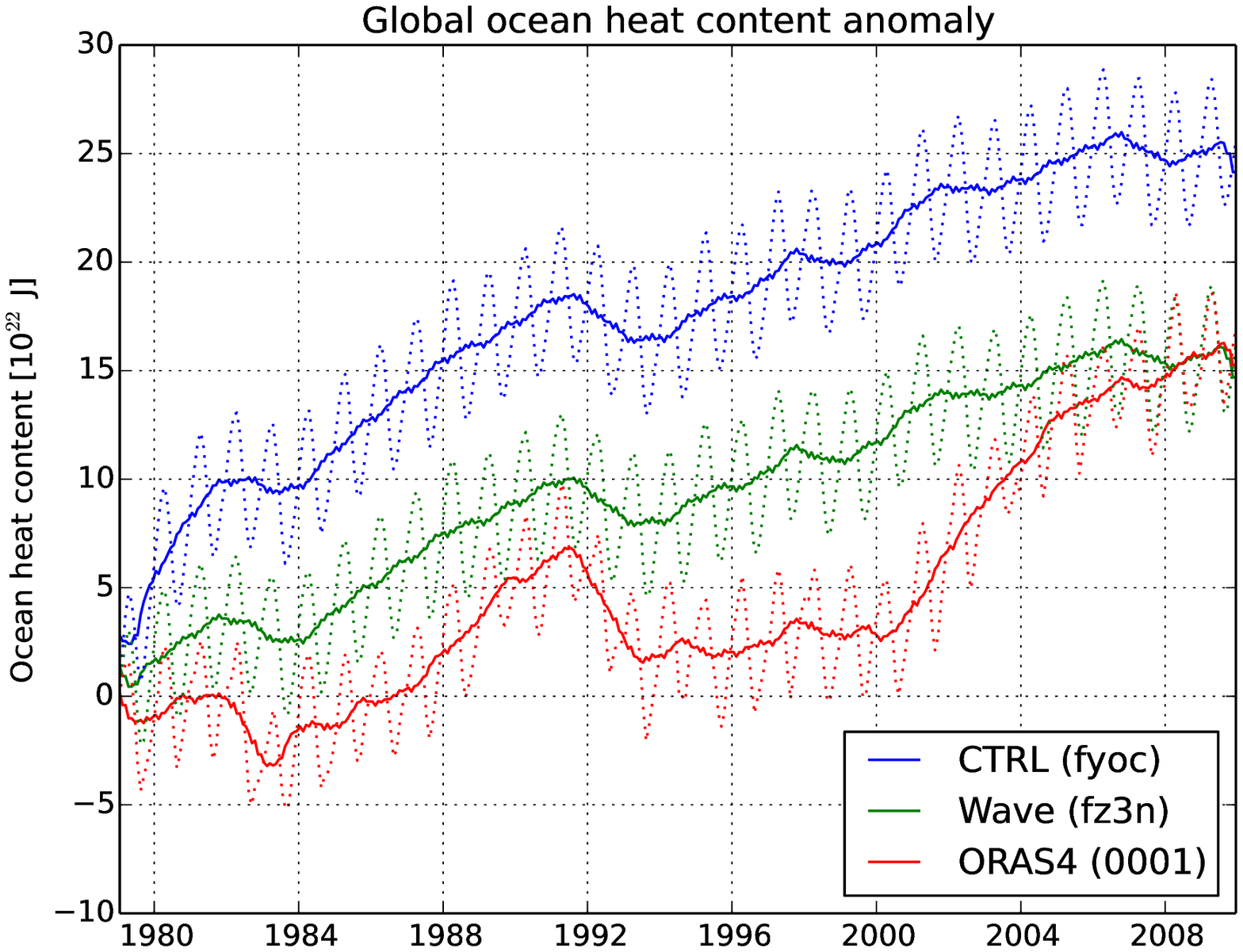}
\end{tabular}
\caption{Panel a: The global ocean heat content anomaly (relative to the start
of the time series) in the upper 300 m.  The run with all three wave effects
switched on is shown in green (see Table 2).  Full lines represent a 12-month
moving average. The ORAS4 reanalysis is shown in red and the CTRL run in blue.
Panel b: Total ocean heat content anomaly.}
\label{fig:ohc}
\end{center}
\end{figure}

We now combine the three wave experiments TAUOC, WTKE and STCOR in one
experiment referred to as the \textbf{WAVE} run.  \Fig{nxtrp} shows the
standard deviation (upper curves) and the biases of the CTRL run (blue) (blue)
(blue) (blue) (blue) (blue) (blue) (blue) and a wave run including all three
wave effects (green). The most striking feature is the large-amplitude annual
cycle observed for the bias of the CTRL run. This amplitude is much weaker for
the run with wave effects, and the seasonality is not at all so clear. Due to
the huge differences to the mixing in the runs with and without wave effects,
the heat uptake of the ocean also differs significantly.  \Fig{ohc} shows the
global total heat content anomaly relative to 1979 (the start of the period)
for the experiment with wave effects (green) v the CTRL experiment (blue).
We also show the ocean reanalysis ORAS4 (red) which is based on an earlier
version of NEMO (v3.0), see \citet{balmaseda13}, with observations assimilated
using the NEMOVAR 3D-VAR assimilation system \citep{mogensen12b}. The
reanalysis covers the period 1957 to present, with forcing provided by ERA-40
\citep{upp05} and later ERA-Interim from 1979 and onwards.  The similarity
with ORAS4 is clear, and the trends almost identical. It is also clear that
the impact of the wave mixing establishes itself within the first two years
of the integration.  The pronounced annual cycle (dashed lines) is due to
the difference in oceanic volume in the southern and northern hemispheres.

\begin{figure}[h]
\begin{center}
\begin{tabular}{cc}
   (a) & \includegraphics[scale=0.75]{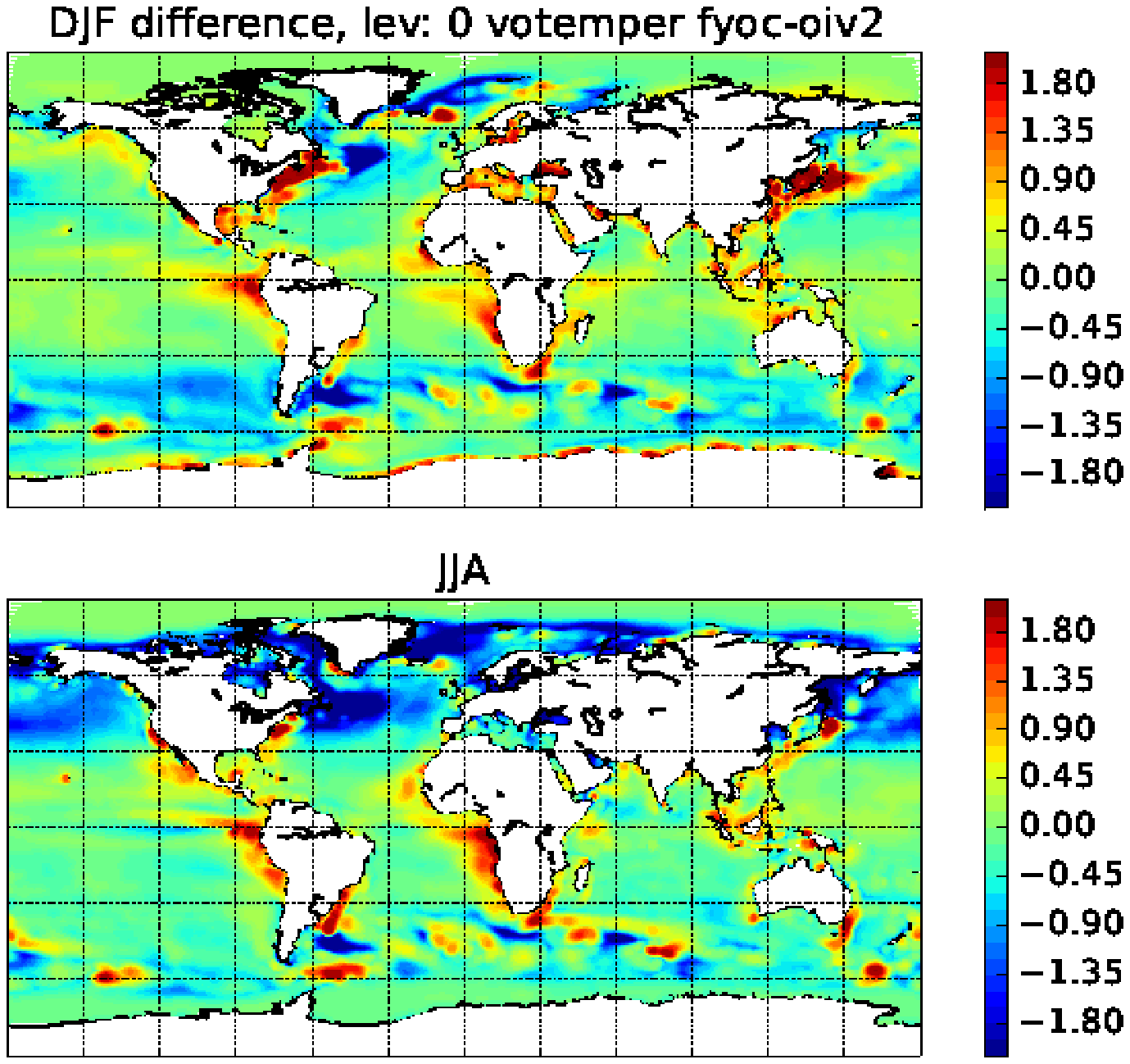}\\ 
   (b) & \includegraphics[scale=0.75]{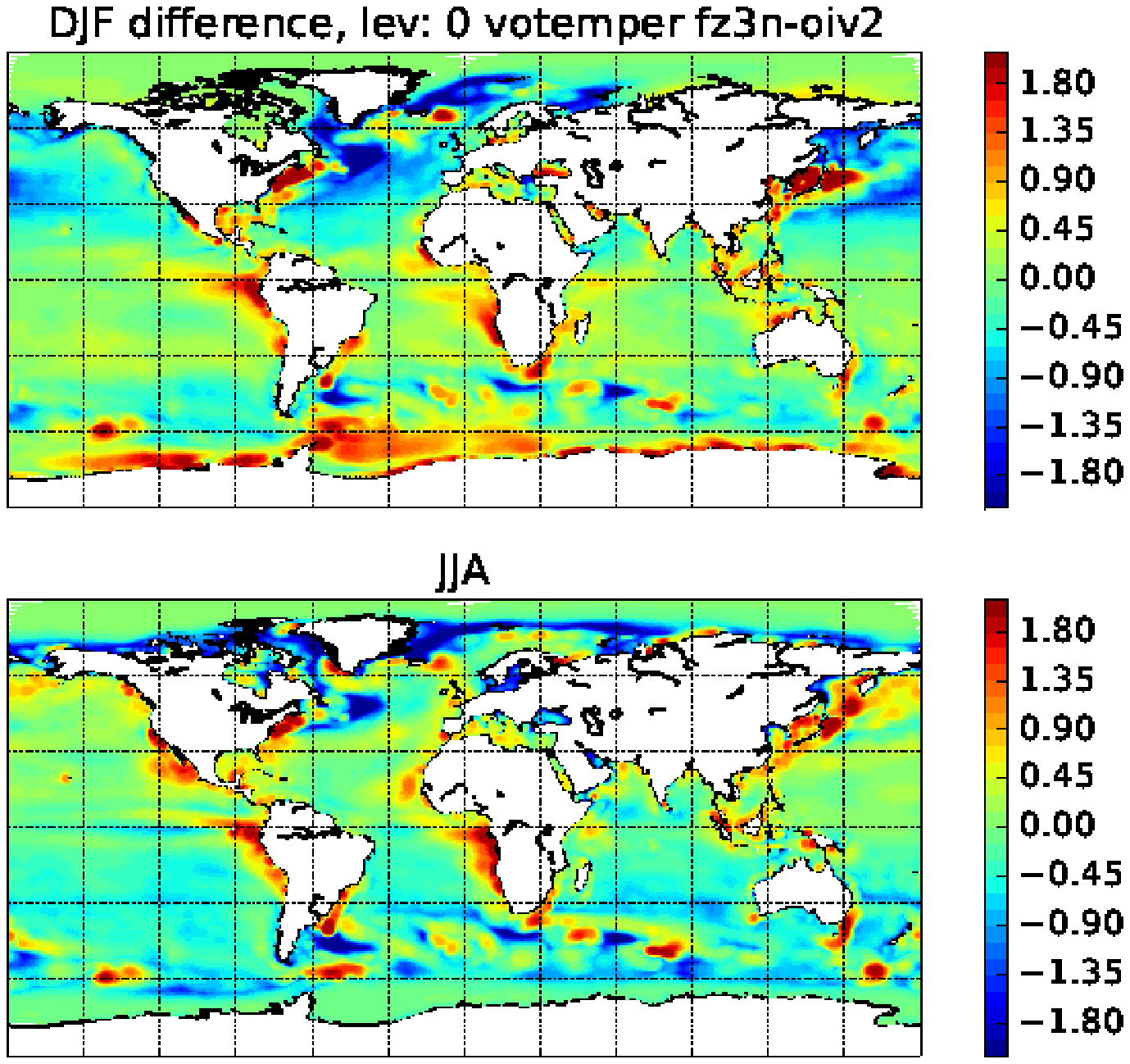}\\
\end{tabular}
\caption{Panel a: Long-term SST differences between the \citet{reynolds02}
OI\_v2 SST analysis and the CTRL run.
Panel b: Differences between the run with wave effects and the CTRL run.}
\label{fig:oiv2}
\end{center}
\end{figure}

To further assess the impact on the surface temperature we now compare with
OI\_v2, an SST analysis by \citet{reynolds02}. The comparison with the CTRL run
in \Fig{oiv2}a reveals large biases, especially in the summer hemisphere. These
biases are reduced when wave effects are included (\Fig{oiv2}b), especially
in the northern extra-tropics. This is mainly due to a more correct level
of mixing.  The long-term SST fields from ORAS4 are very similar to OI\_v2
since a strong relaxation to the OI\_v2 gridded SST was applied as a flux
correction between December 1981 and December 2009 \citep{balmaseda13}
and are not shown here.

\begin{figure}[h]
\begin{center}
\includegraphics[scale=0.8]{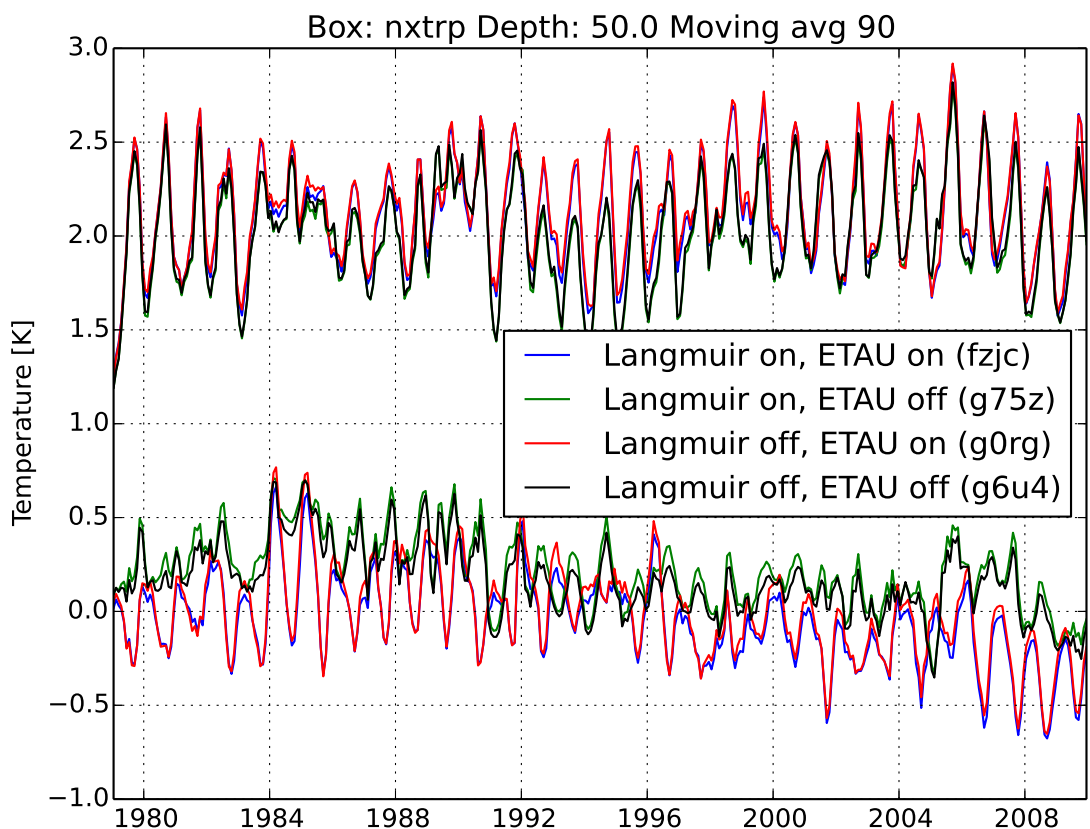}
\caption{A comparison of the impact of ETAU, the parameterization for
enhanced deep mixing in NEMO, and the Langmuir mixing parameterization
by \citet{axell02}.  The northern extra-tropics (defined as north of
20$^\circ$~N) at 50 m depth is shown here.  The blue curve shows the CTRL
run where both processes are switched on (default).  Switching off the
Langmuir mixing (red) is seen to have a much smaller impact than switching
off the ETAU parameterization (green). Finally, a run where both processes
are switched off (black) shows that the combined effect is again dominated by
the ETAU parameterization.  Modeled temperature is compared to EN3 temperature
observations \citep{ingleby07}.  The upper curves show the standard deviation
while the lower curves represent the bias. A 90-day running mean is employed.}
\label{fig:langmuir_etau}
\end{center}
\end{figure}

Finally, to test the relative impact of the ETAU TKE boost (\ref{eq:etau})
and the Langmuir turbulence parameterization (\ref{eq:lc}), we switched
off the Langmuir circulation (\textbf{NOLC}) and ETAU (\textbf{NOETAU})
in two additional experiments.  As is evident from \Fig{langmuir_etau},
the Langmuir turbulence has only modest impact on the temperature in the
mixed layer in the extra-tropics (50 m depth in the northern extra-tropics
is shown in \Fig{langmuir_etau}). Switching off ETAU has a much larger impact
on the mixing, both in terms of bias and random error.

\section{The Coupled Atmosphere-Wave-Ocean Model}
\label{sec:coupled}
\begin{figure}[h]
\begin{center}
\includegraphics[scale=0.9]{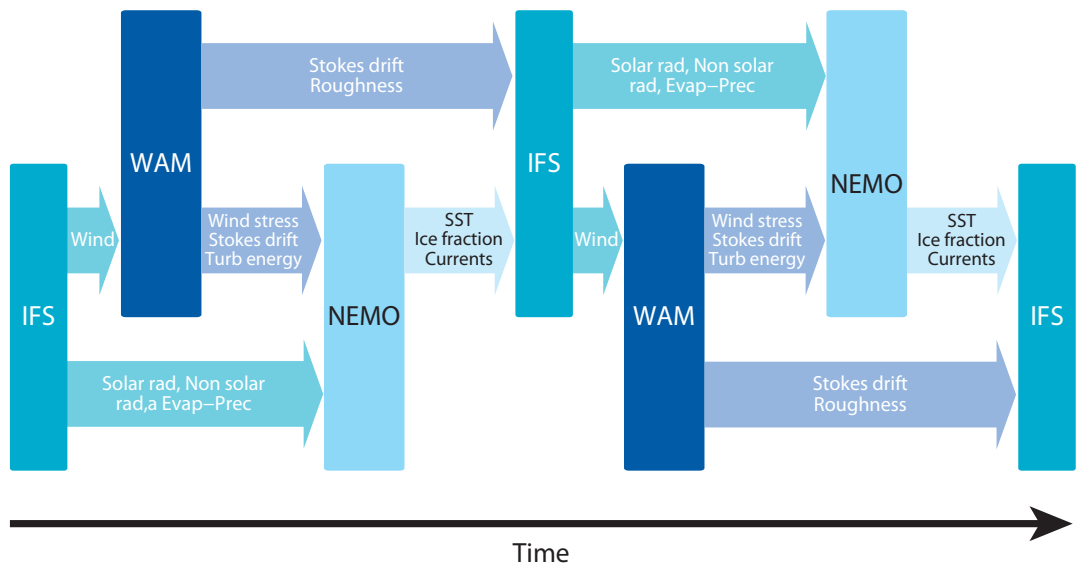}
\caption{A flow chart of the coupling between the components of the single
executable IFS-ECWAM-NEMO model setup. Two coupling time steps are shown
to illustrate the sequence. This is how the operational ensemble prediction
system has been run since Cycle 40R1 (November 2013). The seasonal integration
experiments described here contain in addition the LIM2 ice model and are
run at lower atmospheric resolution (T255).  First the atmospheric component
(IFS) is integrated (internal time step 2,700 s) one coupling time step
(10,800 s), yielding wind fields for ECWAM as well as radiation and evaporation
minus precipitation fields for NEMO.  ECWAM (internal time step 900 s)
is tightly coupled to IFS and gives sea surface roughness for the atmospheric
boundary layer at every atmospheric time step.  After one NEMO coupling time
step, ECWAM also gives Stokes drift velocity, turbulent kinetic energy and
waterside stress to NEMO.  NEMO is integrated (internal time step 3,600 s)
one coupling time step, yielding SST and surface currents to IFS.  All model
components have now been integrated one coupling time step (10,800 s),
and the sequence begins anew with the next coupling time step.}
\label{fig:coupling}
\end{center}
\end{figure}

\begin{figure}[h]
\begin{center}
 (a)\\
 \includegraphics[scale=0.8, bb=308 68 596 555, clip=true, angle=-90]{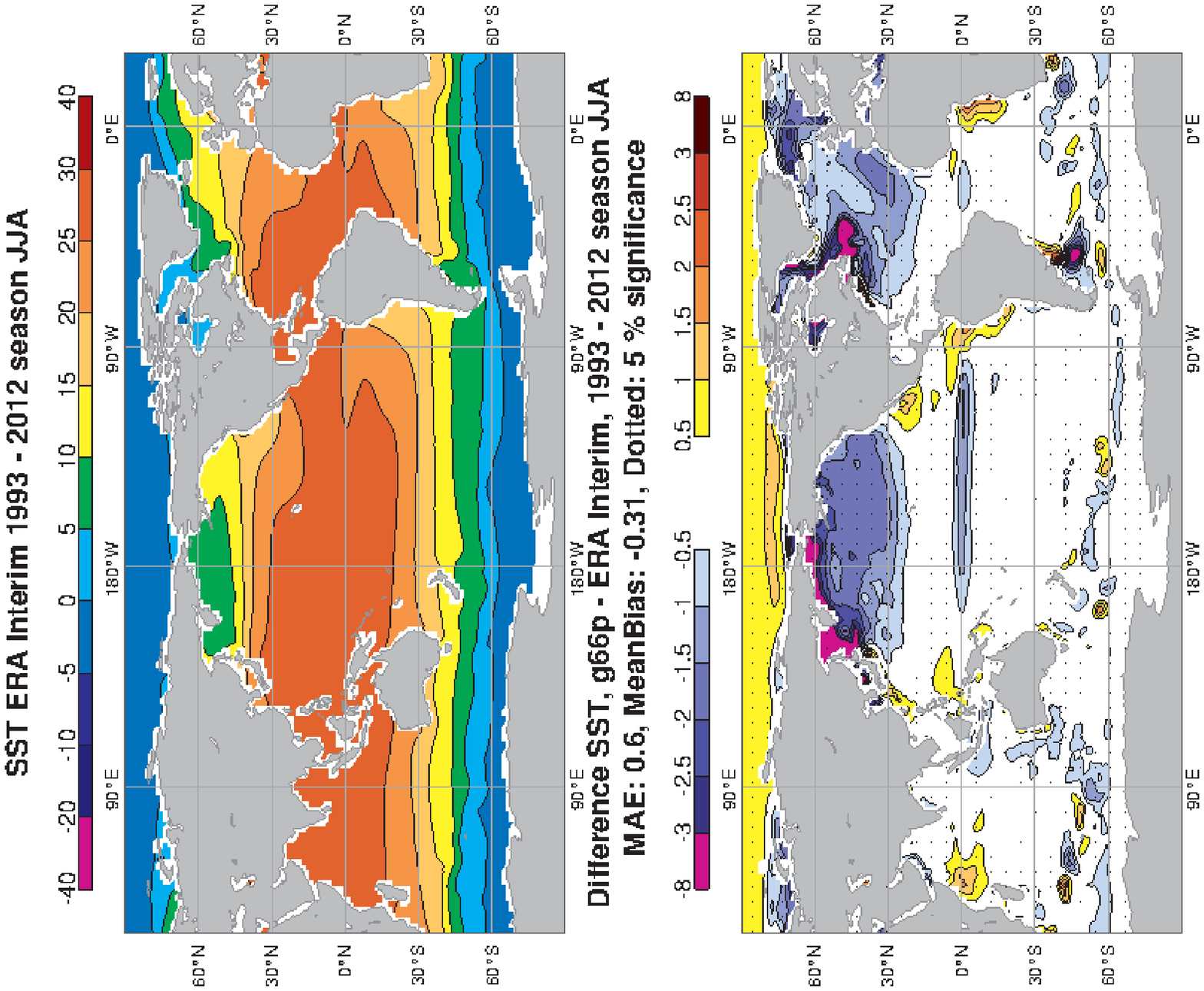}\\
 (b)\\
 \includegraphics[scale=0.8, bb=308 68 596 555, clip=true, angle=-90]{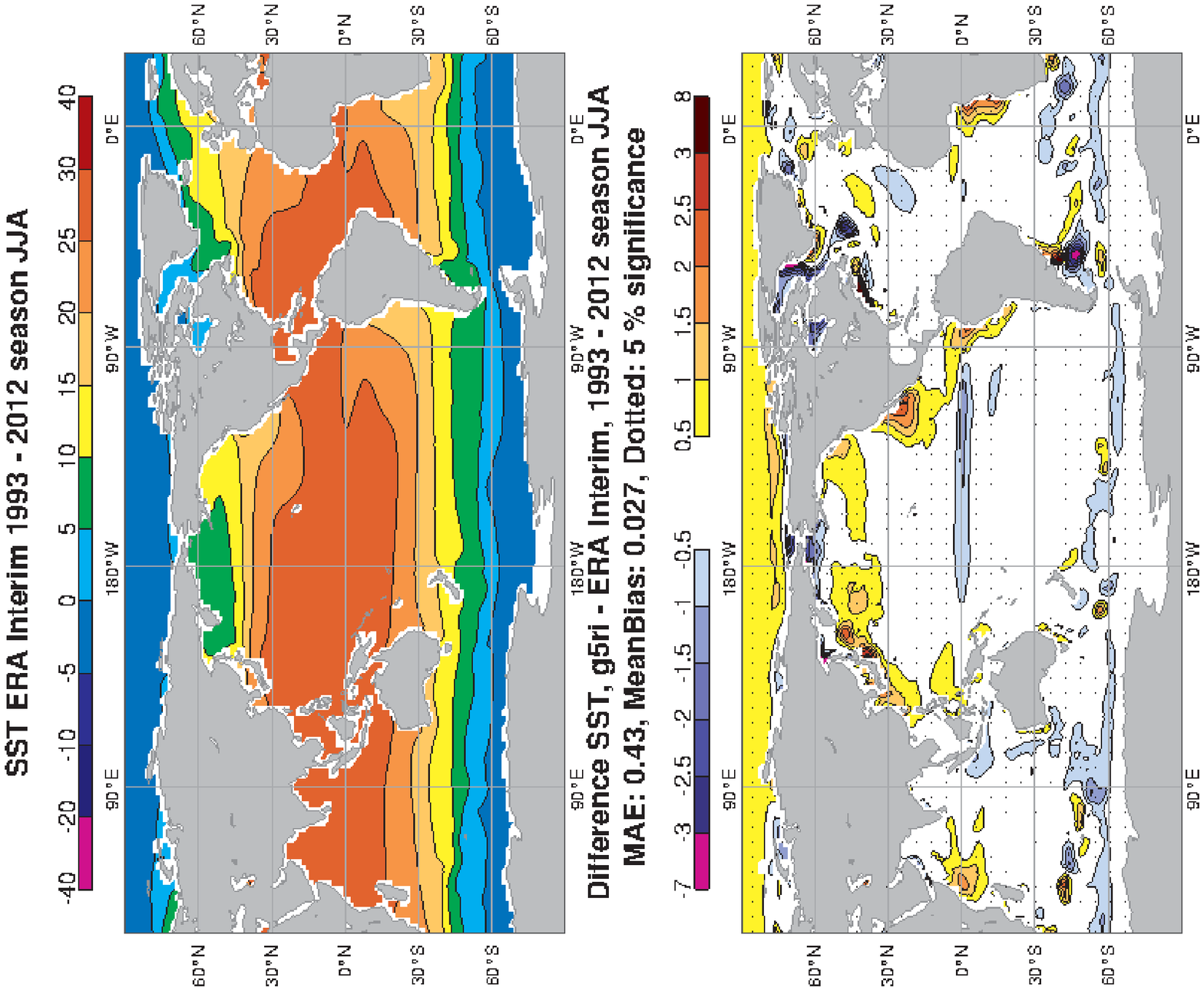}
\caption{Panel a: SST bias relative to ERA-Interim for the coupled seasonal
CTRL run (JJA). Panel b: Same for the wave run.}
\label{fig:sstcoupled}
\end{center}
\end{figure}

\begin{figure}[h]
\begin{center}
  \includegraphics[scale=0.6,angle=-90]{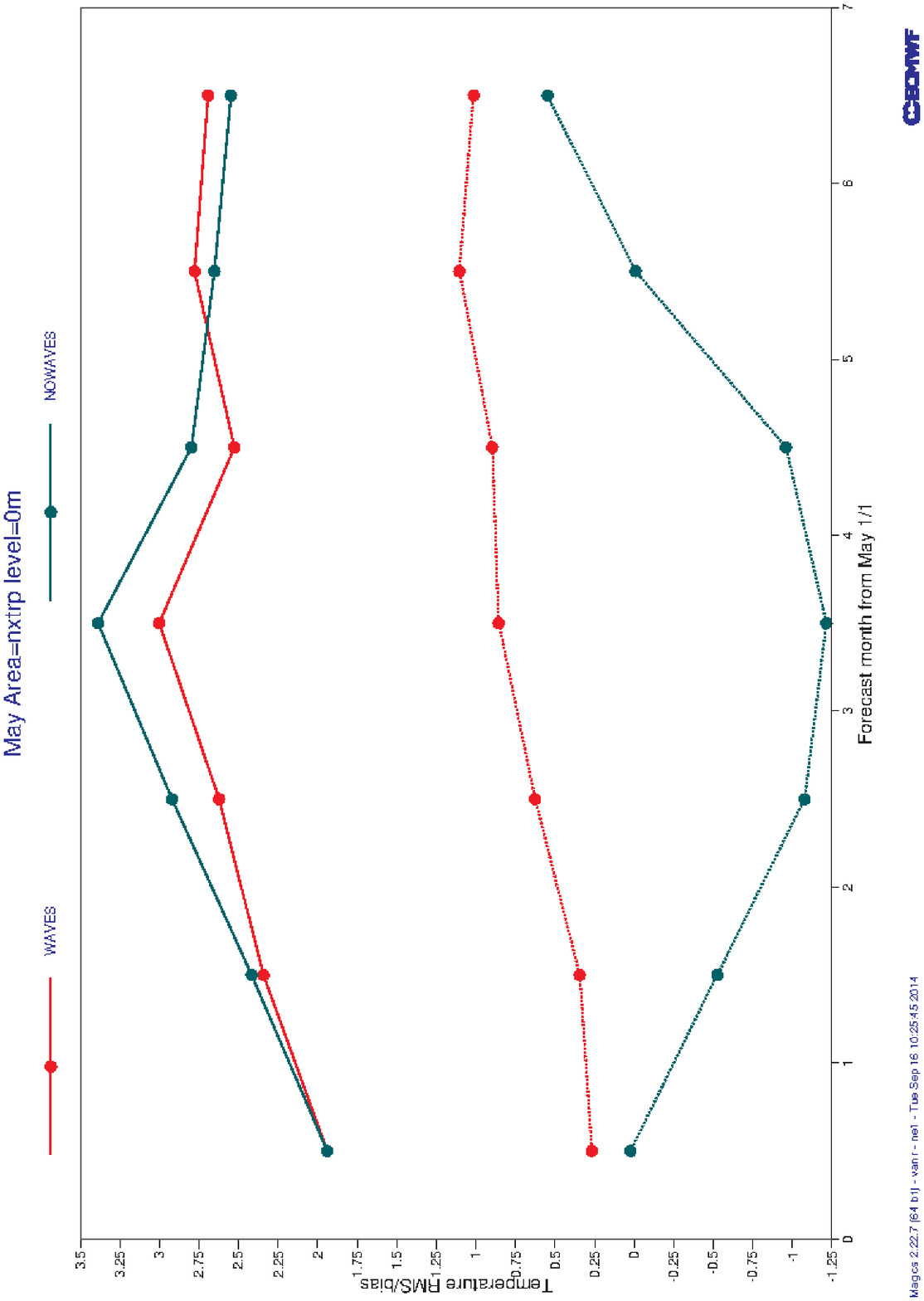}
\caption{A comparison of surface EN3 temperature observations \citep{ingleby07}
in the northern extra-tropics for coupled integrations of IFS-ECWAM-NEMO as
a function of lead time. The start date is 1 May. A three-member ensemble
is run to a lead time of seven months for the years 1993--2008. The dashed
red (lower) curve shows the bias of the run with wave effects, and
the green dashed curve shows the bias of the CTRL run.
The upper full lines show the root-mean-square (RMS) error.}
\label{fig:nxtrpcoupled}
\end{center}
\end{figure}

Coupling between the wave model and the ocean model was first implemented
operationally in the ensemble suite of the IFS in Cycle 40R1 in November
2013, but was limited to the mixing (WTKE) and Stokes-Coriolis forcing (STCOR).
The seasonal integrations described here include in addition the modified
stress (TAUOC) and the ice model LIM2 but are run at lower spatial resolution
than the operational ensemble forecasts. Fields are exchanged every three hours
(the coupling timestep) between IFS/ECWAM and NEMO. The atmospheric model is
run with a spectral truncation of T255 (corresponding to roughly 78 km) with 91
vertical levels to 1 Pa. ECWAM is run at 1.5$^\circ$ resolution while NEMO is
run on the ORCA~1$^\circ$ grid described in \Sec{forced}. \Fig{coupling} shows
a flow chart outlining the sequence of execution of the various components
of the IFS-ECWAM-NEMO single executable over two coupling timesteps, which
can be summarized as follows.
\begin{enumerate}
 \item The atmospheric component (IFS) is integrated (internal time step
 2,700 s) one coupling time step (10,800 s), yielding wind fields for ECWAM
 as well as radiation and evaporation minus precipitation fields for NEMO
 \item ECWAM (internal time step 900 s) is tightly coupled to IFS and returns
 sea surface roughness to the atmospheric boundary layer at every atmospheric
 time step.  After one NEMO coupling time step, ECWAM also gives Stokes
 drift velocity, turbulent kinetic energy and waterside stress to NEMO
 \item NEMO is integrated (internal time step 3,600 s) one coupling time step,
 yielding SST and surface currents to IFS
 \item All model components have now been integrated one coupling time step
 and the sequence begins anew
\end{enumerate}
Surface currents and SST are communicated back to the atmospheric model and
will affect the stress and the temperature of the boundary layer.  This in turn
affects the oceanic wave field, but there is presently no direct feedback from
NEMO to ECWAM.  The coupling between the different components is described
in more detail by \citet{mogensen12}.  Here we compare a setup for seasonal
integrations to seven months, with three ensemble members starting from 1
May and 1 November for the period 1993-2012. The CTRL experiment is run with
a standard CB94 type wave mixing which corresponds to $\alpha_\mathrm{CB}
= 100$ (Eq~\ref{eq:esbc}), similar to the ocean-only CTRL experiment
presented in \Sec{forced}. The wind stress is computed using the ECMWF drag
law (\ref{eq:cdec}) and 10-m wind vectors from IFS.  The wave experiment
includes the three processes TAUOC, WTKE and STCOR described in \Sec{theory}
and \Sec{forced}. \Fig{sstcoupled} reveals large differences in the bias in
the northern extra-tropics relative to ERA-Interim for the boreal summer (JJA)
at a lead time of one to three months from 1 May.  Panel (a) shows the bias
of the CTRL run, with large cold biases in the northern extra-tropics. These
biases are broadly similar to what was found from the ocean-only (forced)
runs presented in \Sec{forced} (cf \Fig{wtke-ctrl}).  Panel (b) reveals the
biases in the run with wave effects to be generally much smaller, although
there is a certain deterioration of the upwelling area along the coast of
Baja California.  Note that the cold bias in the eastern equatorial Pacific
(the ``cold tongue'') is reduced slightly.  The results are similar for the
southern hemisphere summer (DJF), although the bias reduction is not as strong
as for the northern hemisphere. The seasonal variation of the bias found for
the forced (ocean only) runs in \Fig{nxtrp} is also present in the coupled
integrations, see \Fig{nxtrpcoupled}.  Again, the bias is greatly reduced
in the wave run.

\section{Discussion}
\label{sec:disc}
We have introduced three wave effects in NEMO, namely the sea-state
dependent water-side stress, the energy flux from breaking waves and the
Stokes-Coriolis force.  Using ocean-only integrations and experiments with a
coupled system consisting of the atmospheric model IFS, the wave model ECWAM
and NEMO, we demonstrated that the impact of the wave effects is particularly
noticeable in the extra-tropics. Of the three processes, the modification
of the mixing (WTKE) has the largest impact (\Fig{wtke-ctrl}), but as we
discuss below, this is also related to the additional mixing found in NEMO.
The impact of the modified stress (TAUOC) and Stokes-Coriolis (STCOR) is also
significant (on the order of 0.5~K locally, see Figs~\ref{fig:tauoc-ctrl}
and \ref{fig:stcor-ctrl}).  It is also important to note that compared
to the law-of-the-wall experiment (LOW), the WTKE differs by only 0.5~K.
This again suggests that the CTRL experiment has too vigorous mixing.
In ocean-only integrations we see a reduction of the temperature bias in the
mixed layer, particularly in the extra-tropical summer (\Fig{nxtrp}). This
manifests itself in a more realistic oceanic heat uptake (\Fig{ohc}). The
coupled seasonal integrations show a similar reduction in bias (compared
to ERA-Interim, Fig~\ref{fig:sstcoupled}) as the ocean-only wave run (see
Fig~\ref{fig:nxtrpcoupled}). The mixing is strongly influenced by the
ETAU parameterization (\ref{eq:etau}).  It is clear that the TKE scheme
(\ref{eq:tke2}) has too shallow mixing without this parameterization,
and it is also clear that the present parameterization of Langmuir
turbulence (\ref{eq:lc}) does very little due to its vertical structure
which has the distinguishing feature that it will put most if not all of the
enhanced turbulence deep into the mixed layer and nothing near the surface.
It is thus unable to transport down enough heat to make a substantial
difference.  This explains why its impact on the temperature in the OSBL
(\Fig{langmuir_etau}) is so much smaller than that from the ETAU term.
Its vertical profile (\ref{eq:wlc}) is very different from Langmuir
parameterizations involving the shear of the Stokes drift velocity profile
\citep{mcwilliams97,polton07,grant09},
\begin{equation}
   -\overline{\mathbf{u}_\mathrm{H}'w'} \cdot \Dp{\mathbf{v}_{\mathrm{s}}}{z}.
   \label{eq:stokesshear}
\end{equation}
Here, $\mathbf{u}_\mathrm{H}'$ is the horizontal velocity fluctuations and
$w'$ the vertical.  Due to the strong shear of the Stokes drift profile
this would add a larger contribution near the surface.  The ETAU profile
(\ref{eq:etau}) is quite similar to (\ref{eq:stokesshear}) and it appears
to act as a parameterization for Langmuir turbulence with its characteristic
exponential decay with depth (\ref{eq:stokesshear}), or similarly mixing by
non-breaking waves \citep{qiao04,babanin06,huang11}. It facilitates deeper
penetration of mixing from surface processes than what is normally assumed
from breaking waves \citep{grant09,belcher12}.

\section{Concluding remarks and further work}
\label{sec:conc}
The ocean-only integrations and coupled seasonal integrations all suggest
that the right level of mixing is very important for reducing the temperature
bias in the upper part of the ocean and also for the oceanic heat uptake.
An important result is that introducing wave-enhanced mixing must be done in
such a way that the thickness of the uppermost layer is accounted for. This
is done with the present implementation by weighting with the thickness of
the layer and is essential with model configurations with a thick uppermost
level, e.g. ORCA1L42 as discussed here.  This would not be necessary if a
flux boundary condition had been used for the TKE from breaking waves, but
this is not the case in NEMO, which uses the surface boundary condition first
proposed by \citet{mellor04}.  The impact of mixing on SST is clearly shown
in \Fig{oiv2} where runs with and without wave effects are compared with the
OI\_v2 SST analysis.  That the seasonal cycle of the CTRL run is distorted by
too vigorous mixing is clear, and Fig \ref{fig:nxtrp} shows that the annual
cycle in biases extends well below the surface.  The conclusions from the
ocean-only experiments that temperature biases are reduced by introduction of
wave-induced mixing are borne out by the seasonal coupled integrations which
essentially show the same bias reduction (\Fig{sstcoupled}). It is important
to note here, though, that the additional \textit{ad hoc} deep mixing in
NEMO interacts with the surface processes and that without this additional
mixing the model fails to mix deeply enough. We speculate that this mechanism
is really masking Langmuir turbulence or mixing from non-breaking waves.
More work is clearly needed with ocean circulation models and coupled
models to fully answer the question of which mixing processes are dominant
in the OSBL, but it is clear that getting the mixing right is a balancing
act between the right deep mixing and the right mixing near the surface,
and these processes are probably all wave-related.

These results are relevant for assessing the impact surface waves have on
climate projections \citep{babanin09b,fan14}, and a natural next step would
be to investigate the impact of waves on long, decadal to century-wide
integrations (see also the Co-ordinated Ocean-Wave Climate Projections
(COWCLIP) initiative, \citet{hemer12}).  One candidate for forcing ocean-only
integrations would be the recently completed ERA 20th century reanalysis
(ERA-20C, see \citealt{poli13}, \citealt{hersbach13}, \citealt{deboisseson14},
and \citealt{dee14}). This opens up the possibility of running century-long
NEMO integrations with wave effects from a state-of-the-art version of ECWAM
\citep{bidlot12} since all relevant parameters have been archived in the new
reanalysis. For coupled climate projections, the nearest candidate would be
EC-Earth \citep{hazeleger10,hazeleger12} which operates a modified version
of an earlier cycle of IFS. Such experiments would help determining the
importance of waves in the climate system, rather than just the impact of
climate change on the wave climate.

\appendix
\section{The dissipation profile}
The exponential profile (\ref{eq:phimb}) for the balance of
\begin{equation}
  \Dp{}{z}\left(lqS_q \Dp{e}{z}\right) = \frac{q^3}{Bl}
  \label{eq:tkecb}
\end{equation}
assumed by \citet{mellor04} is only valid very near the surface where the
mixing length can be assumed constant = $\kappa z_\mathrm{w}$. CB94 presented
the solution to the more general case where the mixing length is allowed to
vary with depth.  This equation has a power-law solution [cf CB94, Eq (23)],
\begin{equation}
   e(z) = e_0 \left(\frac{z_\mathrm{w}}{z_\mathrm{w} - z}\right)^{2n/3}
   \label{eq:phicb}
\end{equation}
where
\begin{equation}
   n = \left(\frac{3}{S_q\kappa^2 B}\right)^{1/2} = 2.4.
   \label{eq:n}
\end{equation}
This leads to a slightly more complicated expression for
the vertical average (\ref{eq:eavg}),
\begin{equation}
  e_1 = e_0 \frac{z_\mathrm{w}}{L(2n/3-1)} \left[1 -
                       \left(1+\frac{L}{z_\mathrm{w}}\right)^{-2n/3+1}\right].
  \label{eq:e1cb}
\end{equation}
than \Eq{e1}. The consequence is that the mixing penetrates about twice
as deep as in the case where the exponential approximation assumed by
\citet{mellor04}.

\section*{Acknowledgments}
This work has been supported by the European Union project MyWave
(grant FP7-SPACE-2011-284455). This paper is in partial fulfilment of MyWave
deliverables D1.3 and D1.4.  All datasets and model integrations presented in
this study are archived in ECMWF's MARS and ECFS databases. 
For more information on how to access MARS and ECFS, see\\
\underline{\texttt{http://old.ecmwf.int/services/archive/}}.\\
We would like
to thank the two anonymous reviewers for thorough reviews with suggestions
which helped us make the article more succinct and to the point.

\begin{table}[h]
\begin{center}
\begin{tabular}{|l|l|}
\hline
Horizontal grid & ORCA $1^\circ$ \\ \hline
Vertical resolution & 42 levs (10~m top lev) \\ \hline
Time step & 3600 s \\ \hline
Time period & 1979-2009 \\ \hline
Atmospheric forcing & ERA-Interim \\ \hline
Data assimilation & OFF \\ \hline
SST damping  & OFF \\ \hline
3D damping to clim & ON (3 yr Newtonian relaxation) \\ \hline
Bulk parameterization & COARE \\ \hline
Ice model & LIM2 \\ \hline
\end{tabular}
\end{center}
\caption{Overview of the settings common to all forced (ocean only)
experiments.}
\label{tab:commonsettings}
\end{table}

\begin{table}[h]
\begin{center}
\begin{tabular}{|l|l|l|l|l|l|}
\hline
                          & \multicolumn{5}{|c|}{Physical process/parameterization}  \\ \hline  
Experiment description    & Stress & TKE flux & Stokes-Coriolis & Langmuir & ETAU \\ \hline \hline
CTRL:                     & Drag law  & CB94 TKE flux & Off & On & On \\ 
Control experiment        & \Eq{cdec} & $\alpha_\mathrm{CB}=100$ \Eq{alphacb} & & & \\ \hline
TAUOC: Water-side         & \textbf{ECWAM} & as CTRL & Off & On & On \\ 
           stress         &  \textbf{stress} (\ref{eq:tauoc})   &         &     &    &     \\ \hline
WTKE: Sea-state           & as CTRL & \textbf{ECWAM} & Off & On & On \\ 
dependent TKE flux        &         &  \textbf{TKE flux} (\ref{eq:phioc})     &  & & \\ \hline
STCOR: Stokes-Coriolis    & as CTRL & as CTRL & \textbf{ECWAM} & On & On \\
 forcing                  &         &         &  \textbf{Stokes} (\ref{eq:stcor}) & &  \\ \hline
WAVE: & \textbf{as TAUOC} & \textbf{as WTKE} & \textbf{as STCOR} & On & On \\ \hline
All three wave effects    &         &   &  &    & \\ \hline
LOW: Law-of-the-wall      & as CTRL & \textbf{Off} & Off & On & On \\ \hline
NOLC: Langmuir off        & as CTRL & as CTRL & Off & \textbf{Off} & On \\ \hline
NOETAU: ETAU off          & as CTRL & as CTRL & Off & On & \textbf{Off} \\ \hline
\end{tabular}
\end{center}
\caption{Overview of the settings of the ocean-only (forced) experiments.
Departures from the CTRL experiment marked with bold.}
\label{tab:settings}
\end{table}

\end{document}